\documentclass[fleqn]{article}
\usepackage{amsmath}
\usepackage[dvips]{graphicx}

\begin{document}

\title{\bf A new integrable generalization\\ of the Korteweg--de~Vries equation}

\author{\sc Ay\c{s}e Karasu-Kalkanl\i\,$^{1)}$\!, Atalay Karasu\,$^{2)}$\!,\\
\sc Anton Sakovich\,$^{3)}$\!, Sergei Sakovich\,$^{4)}$\!, Refik Turhan\,$^{5)}$\\[12pt]
{\footnotesize $^{1,2)}$Department of Physics, Middle East Technical University, 06531 Ankara, Turkey}\\[-4pt]
{\footnotesize $^{3)}$National Centre of Particle and High Energy Physics, 220040 Minsk, Belarus}\\[-4pt]
{\footnotesize $^{4)}$Institute of Physics, National Academy of Sciences, 220072 Minsk, Belarus}\\[-4pt]
{\footnotesize $^{5)}$Department of Engineering Physics, Ankara University, 06500 Ankara, Turkey}\\
{\footnotesize E-mail: $^{1)}$akarasu@metu.edu.tr, $^{2)}$karasu@metu.edu.tr,}\\[-4pt]
{\footnotesize $^{3)}$ant.s@tut.by, $^{4)}$saks@tut.by, $^{5)}$turhan@eng.ankara.edu.tr}}

\date{}

\maketitle

\begin{abstract}
A new integrable sixth-order nonlinear wave equation is discovered by means of the Painlev\'{e} analysis, which is equivalent to the Korteweg--de~Vries equation with a source. A Lax representation and a B\"{a}cklund self-transformation are found of the new equation, and its travelling wave solutions and generalized symmetries are studied.
\end{abstract}

\section{Introduction}

In this paper we apply the Painlev\'{e} test for integrability of partial differential equations \cite{WTC,T} to the class of sixth-order nonlinear wave equations
\begin{multline} \label{e1}
u_{xxxxxx} + a u_x u_{xxxx} + b u_{xx} u_{xxx} + c u_x^2 u_{xx} \\ + d u_{tt} + e u_{xxxt} + f u_x u_{xt} + g u_t u_{xx} = 0 ,
\end{multline}
where $a,b,c,d,e,f$ and $g$ are arbitrary parameters. We show that there are four distinct cases of relations between the parameters when equation \eqref{e1} passes the Painlev\'{e} test well. Three of those cases correspond to known integrable equations, whereas the fourth one turns out to be new. This new integrable case of equation \eqref{e1} is equivalent to the Korteweg--de~Vries equation with a source of a new type, and we find its  Lax pair, B\"{a}cklund self-transformation, travelling wave solutions and third-order generalized symmetries.

There are the following reasons to explore the class of equations \eqref{e1} for integrability. Recently Dye and Parker \cite{DP} constructed and studied two integrable nonlinear integro-differential equations,
\begin{multline} \label{e2}
5 \partial_x^{-1} v_{tt} + 5 v_{xxt} - 15 v v_t - 15 v_x \partial_x^{-1} v_t \\ - 45 v^2 v_x + 15 v_x v_{xx} + 15 v v_{xxx} - v_{xxxxx} = 0
\end{multline}
and
\begin{multline} \label{e3}
5 \partial_x^{-1} v_{tt} + 5 v_{xxt} - 15 v v_t - 15 v_x \partial_x^{-1} v_t \\ - 45 v^2 v_x + \tfrac{75}{2} v_x v_{xx} + 15 v v_{xxx} - v_{xxxxx} = 0 ,
\end{multline}
which describe the propagation of waves in two opposite directions and represent bidirectional versions of the Sawada--Kotera--Caudrey--Dodd--Gibbon equation \cite{SK,CDG} and Kaup--Kupershmidt equation \cite{K,FG}, respectively. Equations \eqref{e2} and \eqref{e3} possess Lax pairs due to their construction \cite{DP} and fall into class \eqref{e1} after the potential transformation $v = u_x$. There is one more well-known integrable equation in class \eqref{e1}, namely
\begin{equation} \label{e4}
	u_{tt} - u_{xxxt} - 2 u_{xxxxxx} + 18 u_x u_{xxxx} + 36 u_{xx} u_{xxx} - 36 u_x^2 u_{xx} =0 .
\end{equation}
This equation is equivalent to the Drinfel'd--Sokolov--Satsuma--Hirota system of coupled Korteweg--de~Vries equations \cite{DS,SH}, of which a fourth-order recursion operator was found in \cite{GK}. A B\"{a}cklund self-transformation of equation \eqref{e4} was derived in \cite{KS} by the method of truncated singular expansion \cite{WTC,W1}. Multisoliton solutions of equations \eqref{e3} and \eqref{e4} were studied in \cite{VM1,VM2,V}. Thus we have already known three interesting integrable equations of class \eqref{e1}, and it is natural to ask what are other integrable equations in this class and, if there are any, what are their properties. Solving problems of this kind is important for testing the reliability of integrability criteria and for discovering new interesting objects of soliton theory.

The paper is organized as follows. In section~\ref{s2} we perform the singularity analysis of equation \eqref{e1} and find four distinct cases which possess the Painlev\'{e} property and correspond, up to scale transformations of variables, to equations \eqref{e2}--\eqref{e4} and to the new integrable equation
\begin{equation} \label{e5}
	\left( \partial_x^3 + 8 u_x \partial_x + 4 u_{xx} \right) \left( u_t + u_{xxx} + 6 u_x^2 \right) = 0 .
\end{equation}
In this part the present study is similar to the recent Painlev\'{e} classifications done in \cite{KK,S1,S2,ST}, where new integrable nonlinear wave equations were discovered as well. The method of truncated singular expansion is successfully used in section~\ref{s3}, where we derive a Lax pair for the new equation \eqref{e5} and also obtain and study its B\"{a}cklund self-transformation. The contents of section~\ref{s4} is not related to the Painlev\'{e} property: there we find and discuss travelling wave solutions and third-order generalized symmetries of equation \eqref{e5}. Section~\ref{s5} contains concluding remarks.

\section{Singularity analysis} \label{s2}

In order to select integrable cases of equation \eqref{e1} we use the so-called Weiss--Kruskal algorithm for singularity analysis of partial differential equations \cite{RGB}, which is based on the Weiss--Tabor--Carnevale expansions of solutions near movable singularity manifolds \cite{WTC}, Ward's requirement not to examine singularities of solutions at characteristics of equations \cite{W2,W3} and Kruskal's simplifying representation for singularity manifolds \cite{JKM}, and which follows step by step the Ablowitz--Ramani--Segur algorithm for ordinary differential equations \cite{ARS}. Computations are made using the Mathematica computer algebra system \cite{W4}, and we omit inessential details.

Equation \eqref{e1} is a sixth-order normal system, and its general solution must contain six arbitrary functions of one variable \cite{O}. A hypersurface $\phi (x,t) = 0$ is noncharacteristic for equation \eqref{e1} if $\phi_x \neq 0$, and we set $\phi_x = 1$ without loss of generality. Substitution of the expansion $u = u_0 (t) \phi^{\delta} + \dotsb + u_r (t) \phi^{r + \delta} + \dotsb$ to equation \eqref{e1} determines branches of the dominant behavior of solutions near $\phi = 0$, i.e.\ admissible choices of $\delta$ and $u_0$, and corresponding positions $r$ of the resonances, where arbitrary functions of $t$ can enter the expansion. There are two singular branches, both with $\delta = -1$, values of $u_0$ being the roots of a quadratic equation with constant coefficients. Without loss of generality we make $u_0 = 1$ for one of the two branches by a scale transformation of $u$, thus fixing the coefficient $c$ of equation \eqref{e1} as
\begin{equation} \label{e6}
	c = 12 a + 6 b - 360 ,
\end{equation}
and then we get $u_0 = 60 / (2 a + b - 60)$ for the other branch. We require that at least one of singular branches is a generic one, representing the general solution of equation \eqref{e1}, and without loss of generality we assume that we have set $u_0 = 1$ for the generic branch, whereas the branch with  $u_0 = 60 / (2 a + b - 60)$ may be nongeneric. Positions $r$ of the resonances are determined by the equation
\begin{equation} \label{e7}
	(r+1)(r-1)(r-6) \left( r^3 - 15r^2 + (86-a)r + (4a+2b-240) \right) = 0
\end{equation}
for the branch with $u_0 = 1$, and by the equation
\begin{multline} \label{e8}
	(r+1)(r-1)(r-6) \Bigl( r^3 - 15r^2 + \bigl( 86 - 60a / (2a+b-60) \bigr) r \\ + \bigl( -120 + 7200 / (2a+b-60) \bigr) \Bigr) = 0
\end{multline}
for the branch with $u_0 = 60 / (2 a + b - 60)$.

Let us consider the generic branch with $u_0 = 1$ first. From equation \eqref{e7} we find that positions of three resonances are $r = -1, 1, 6$, where $r = -1$ corresponds to the arbitrary dependence of $\phi$ on $t$, and that the positions $r_1$, $r_2$ and $r_3$ of other three resonances satisfy the relations
\begin{gather}
	r_3 = 15 - r_1 - r_2 , \label{e9} \\
	a = r_1^2 + r_1 r_2 + r_2^2 - 15 r_1 - 15 r_2 + 86 , \label{e10} \\
	b = \tfrac{1}{2} r_1^2 r_2 + \tfrac{1}{2} r_1 r_2^2 - 2 r_1^2 - \tfrac{19}{2} r_1 r_2 - 2 r_2^2 + 30 r_1 + 30 r_2 - 52 . \label{e11}
\end{gather}
In order to have the resonances of this generic branch in admissible positions, we must set the numbers $r_1$, $r_2$ and $r_3$ to be positive, integer, distinct and not equal to $1$ or $6$. Assuming without loss of generality that $r_1 < r_2 < r_3$ and taking relation \eqref{e9} into account, we get five distinct cases to be studied further: (i) $r_1 = 2$, $r_2 = 3$; (ii) $r_1 = 2$, $r_2 = 4$; (iii) $r_1 = 2$, $r_2 = 5$; (iv) $r_1 = 3$, $r_2 = 4$; (v) $r_1 = 3$, $r_2 = 5$.

{\sc Case (i):} $r_1 = 2$, $r_2 = 3$. From relations \eqref{e9}--\eqref{e11} and \eqref{e6} we find that $r_3 = 10$, $a = 30$, $b = 30$ and $c = 180$. In the generic branch we have $u_0 = 1$ and $r = -1, 1, 2, 3, 6, 10$. Substitution of the expansion $u = \sum_{i=0}^{\infty} u_i (t) \phi^{i-1}$ with $\phi_x (x,t) = 1$ to equation \eqref{e1} determines recursion relations for $u_i$, and we check whether those recursion relations are compatible at the resonances. We must set $e = \tfrac{1}{12} ( f + g )$ for the compatibility condition at $r = 2$ to be satisfied identically. In the same way we get the relations $f = g$ and $d = - \tfrac{1}{180} g^2$ at the resonances $r = 6$ and $r = 10$, respectively. With the obtained constraints on the coefficients of equation \eqref{e1} we proceed to the nongeneric branch, where $u_0 = 2$ and $r = -2, -1, 1, 5, 6, 12$, and find that the recursion relations are compatible there, which completes the Painlev\'{e} test. Finally, it is easy to prove that, under the obtained constraints on the coefficients, a scale transformation of variables relates equation \eqref{e1} to the known integrable equation \eqref{e2} with $v = u_x$.

{\sc Case (ii):} $r_1 = 2$, $r_2 = 4$. This case of equation \eqref{e1} with $a = 24$, $b = 36$, $c = 144$ and $r_3 = 9$ does not pass the Painlev\'{e} test for integrability. We find that in the nongeneric branch, where $u_0 = \tfrac{5}{2}$, three of the resonances lie in the noninteger positions $r \approx -2.54656, \, 6.26589, \, 11.2807$ due to equation \eqref{e8}, these positions being the irrational roots of $r^3 - 15 r^2 + 26 r + 180 = 0$. Moreover, in the generic branch, where $u_0 = 1$ and $r = -1, 1, 2, 4, 6, 9$, the recursion relations fail to be compatible at the resonance $r = 4$, and this means that the general solution of equation \eqref{e1} contains nondominant movable logarithmic singularities.

{\sc Case (iii):} $r_1 = 2$, $r_2 = 5$. Here we have $r_3 = 8$, $a = 20$, $b = 40$ and $c = 120$ due to relations \eqref{e9}--\eqref{e11} and \eqref{e6}. In the generic branch, where $u_0 = 1$ and $r = -1, 1, 2, 5, 6, 8$, we get the constraints $e = \tfrac{1}{12} ( f + g )$, $f = 2 g$ and $d = 0$ at the resonances $r = 2$, $r = 5$ and $r = 8$, respectively. Under these constraints on the coefficients of equation \eqref{e1}, the recursion relations are compatible in the nongeneric branch as well, where $u_0 = 3$ and $r = -3, -1, 1, 6, 8, 10$. Finally, a scale transformation of variables leads us to the new equation \eqref{e5}, which must be integrable according to the obtained result of the Painlev\'{e} test.

{\sc Case (iv):} $r_1 = 3$, $r_2 = 4$. We find that $r_3 = 8$, $a = 18$, $b = 36$ and $c = 72$. In the generic branch, where $u_0 = 1$ and $r = -1, 1, 3, 4, 6, 8$, we get $g = 0$ and $d = -2 e^2 + \tfrac{1}{2} e f - \tfrac{1}{36} f^2$ at the resonance $r = 4$, and $f = 0$ at the resonance $r = 6$. No extra constraints on the coefficients of equation \eqref{e1} arise in the nongeneric branch, where $u_0 = 5$ and $r = -5, -1, 1, 6, 8, 12$, and we obtain the known integrable equation \eqref{e4} after a scale transformation of variables.

{\sc Case (v):} $r_1 = 3$, $r_2 = 5$. We find that $r_3 = 7$, $a = 15$, $b = \tfrac{75}{2}$ and $c = 45$. In the generic branch, where $u_0 = 1$ and $r = -1, 1, 3, 5, 6, 7$, we get $d = \tfrac{1}{90} ( f^2 - f g - 2 g^2 )$ and $e = \tfrac{1}{6} ( f + g )$ at $r = 5$, and $f = g$ at $r = 6$. No extra constraints on the coefficients of equation \eqref{e1} arise in the nongeneric branch, where $u_0 = 8$ and $r = -7, -1, 1, 6, 10, 12$, and a scale transformation of variables leads us to the known integrable equation \eqref{e3} with $v = u_x$.

We have completed the Painlev\'{e} analysis of equation \eqref{e1}. It is noteworthy that the Painlev\'{e} test not only detected all the previously known integrable cases \eqref{e2}--\eqref{e4} of equation \eqref{e1}, but also discovered the new case \eqref{e5} which is integrable due to the results of the next section. In what follows we refer to equation \eqref{e5} as the KdV6, in order to emphasize that this new integrable sixth-order nonlinear wave equation is associated with the same spectral problem as of the potential Korteweg--de~Vries equation. Let us also note that the KdV6, i.e.\ equation \eqref{e5}, is equivalent to the following Korteweg--de~Vries equation with a source satisfying a third-order ordinary differential equation:
\begin{equation} \label{e12}
	v_t + v_{xxx} + 12 v v_x - w_x = 0 , \qquad w_{xxx} + 8 v w_x + 4 w v_x = 0 ,
\end{equation}
where the new dependent variables $v$ and $w$ are related to $u$ as
\begin{equation} \label{e13}
	v = u_x , \qquad w = u_t + u_{xxx} + 6 u_x^2 .
\end{equation}
The system of equations \eqref{e12} is different from the so-called Korteweg--de~Vries equation with a self-consistent source which was extensively studied during last decade (see \cite{LYZ} and references therein).

\section{Truncated singular expansion} \label{s3}

Let us try to find a Lax pair and a B\"{a}cklund self-transformation for the KdV6, using the method of truncated singular expansion \cite{WTC,W1}.

We substitute the truncated singular expansion
\begin{equation} \label{e14}
	u = \dfrac{y(x,t)}{\phi (x,t)} + z(x,t)
\end{equation}
to equation \eqref{e5} (note that the Kruskal simplifying representation $\phi_x = 1$ is not used now), collect terms with equal degrees of $\phi$, and in this way get the following. At $\phi^{-7}$ we have three possibilities: $y = 0$, $y = \phi_x$ or $y = 3 \phi_x$. We choose
\begin{equation} \label{e15}
	y = \phi_x
\end{equation}
which corresponds to the generic branch, i.e.\ we truncate the singular expansion representing the general solution of the KdV6. Then we get identities at $\phi^{-6}$ and $\phi^{-5}$, whereas the terms with $\phi^{-4}$ give us the equation
\begin{equation} \label{e16}
	z_{xx} = - \dfrac{\phi_{xxxx}}{4 \phi_x} + \dfrac{\phi_{xx} \phi_{xxx}}{2 \phi_x^2} - \dfrac{\phi_{xx}^3}{4 \phi_x^3}
\end{equation}
which is equivalent to
\begin{equation} \label{e17}
	z_{x} = - \dfrac{\phi_{xxx}}{4 \phi_x} + \dfrac{\phi_{xx}^2}{8 \phi_x^2} + \sigma ,
\end{equation}
where $\sigma = \sigma (t)$ is an arbitrary function that appeared as the `constant' of integration of equation \eqref{e16} over $x$. Next, using the obtained relations \eqref{e15} and \eqref{e17}, we get at $\phi^{-3}$ the equation
\begin{multline} \label{e18}
	z_t = - \dfrac{\phi_{xxxxx}}{4 \phi_x} + \dfrac{5 \phi_{xx} \phi_{xxxx}}{4 \phi_x^2} + \dfrac{5 \phi_{xxx}^2}{8 \phi_x^2} - \dfrac{25 \phi_{xx}^2 \phi_{xxx}}{8 \phi_x^3} + \dfrac{45 \phi_{xx}^4}{32 \phi_x^4} \\ - \dfrac{\phi_{xxt}}{2 \phi_x} + \dfrac{\phi_{xx} \phi_{xt}}{2 \phi_x^2} + \sigma \left( - \dfrac{5 \phi_{xxx}}{\phi_x} + \dfrac{15 \phi_{xx}^2}{2 \phi_x^2} - \dfrac{2 \phi_t}{\phi_x} \right) - 30 \sigma^2 .
\end{multline}
The terms with $\phi^{-2}$ and $\phi^{-1}$ add nothing to the already obtained relations \eqref{e15}, \eqref{e17} and \eqref{e18}. Finally, at $\phi^0$ we find that $\sigma '(t) = 0$, i.e.\ $\sigma$ is an arbitrary constant in relations \eqref{e17} and \eqref{e18}.

We have found that the truncation procedure is consistent for the KdV6. The function $u$ determined by relations \eqref{e14}, \eqref{e15}, \eqref{e17} and \eqref{e18} is a solution of equation \eqref{e5}, as well as the function $z$ determined by relations \eqref{e17} and \eqref{e18} is. Moreover, these expressions for $u$ and $z$ correspond to the general solution of the KdV6, because the function $\phi$ is determined by a sixth-order equation---the compatibility condition $z_{xt} = z_{tx}$ for equations \eqref{e17} and \eqref{e18}---and the general solution for $\phi$ contains six arbitrary functions of one variable.

We can derive a Lax pair for the KdV6 from the system of equations \eqref{e17} and \eqref{e18} in the same way as used for the Korteweg--de~Vries equation in \cite{WTC}. Introducing the function $\psi$ related to $\phi$ as
\begin{equation} \label{e19}
	\phi_x = \psi^2 ,
\end{equation}
we immediately get equation \eqref{e17} linearized:
\begin{equation} \label{e20}
	\psi_{xx} + 2 ( z_x - \sigma) \psi = 0 .
\end{equation}
Then we multiply equation \eqref{e18} by $\phi_x$, apply $\partial_x$ to the result, eliminate all derivatives of $\phi$ using relation \eqref{e19}, eliminate derivatives of $\psi$ of order higher than one using equation \eqref{e20}, and in this way obtain the equation
\begin{multline} \label{e21}
	\psi_t = \Bigl( - 8 \sigma - 4 z_x - \tfrac{1}{2} \sigma^{-1} \left( z_t + z_{xxx} + 6 z_x^2 \right) \Bigr) \psi_x \\ + \Bigl( 2 z_{xx} + \tfrac{1}{4} \sigma^{-1} \left( z_{xt} + z_{xxxx} + 12 z_x z_{xx} \right) \Bigr) \psi ,
\end{multline}
also linear with respect to $\psi$. Equations \eqref{e20} and \eqref{e21} are compatible for $\psi$ if and only if $z$ is a solution of equation \eqref{e5}. Consequently, equations \eqref{e20} and \eqref{e21}, where $z$ should be re-denoted as $u$, constitute a Lax pair for equation \eqref{e5}, the arbitrary constant $\sigma$ being a spectral parameter. Note that the spectral problem \eqref{e20}, with $z$ re-denoted as $u$, is the same of equation \eqref{e5} and of the potential Korteweg--de~Vries equation
\begin{equation} \label{e22}
	u_t + u_{xxx} + 6 u_x^2 = 0 .
\end{equation}
Note also that in variables \eqref{e13} the obtained Lax pair of the KdV6 turns into the Lax pair of the Korteweg--de~Vries equation with a source  \eqref{e12}:
\begin{gather}
	\psi_{xx} + 2 ( v - \sigma) \psi = 0, \label{e23} \\ \psi_t = \left( - 8 \sigma - 4 v - \tfrac{1}{2} \sigma^{-1} w \right) \psi_x + \left( 2 v_x + \tfrac{1}{4} \sigma^{-1} w_x \right) \psi . \label{e24}
\end{gather}

From the truncated singular expansion we can also derive a B\"{a}cklund self-transformation for the KdV6. We introduce the notations
\begin{equation} \label{e25}
	p = u - z , \qquad q = u + z ,
\end{equation}
eliminate $\phi$ from relations \eqref{e14}, \eqref{e15}, \eqref{e17} and \eqref{e18}, and in this way obtain the following two equations for $p$ and $q$:
\begin{equation} \label{e26}
	p_{xx} - \tfrac{1}{2} p^{-1} p_x^2 + \tfrac{1}{2} p \left( 4 q_x + p^2 - 8 \sigma \right) =0
\end{equation}
and
\begin{multline} \label{e27}
	q_{xxxx} - p^{-1} p_x \left( q_{xxx} + 3 q_x^2 + q_t + 8 \sigma q_x + 32 \sigma^2 \right) + 2 \left( 3 q_x - p^2 + 4 \sigma \right) q_{xx} \\ + q_{xt} - 5 p p_x \left( 2 q_x + p^2 - 8 \sigma \right) + p p_t - 4 \sigma p^{-1} p_t = 0 ,
\end{multline}
where $\sigma$ is an arbitrary constant. These equations \eqref{e26} and \eqref{e27} constitute a B\"{a}cklund transformation according to the definition used in \cite{AS}. Namely, when we eliminate $z$ from equations \eqref{e26} and \eqref{e27}, we exactly get equation \eqref{e5} for $u$; and vice versa, we get equation \eqref{e5} for $z$ if we eliminate $u$ from system \eqref{e26}--\eqref{e27}. However, one definitely needs to use computer algebra tools to prove this.

Some words are due on the following interesting property of the obtained B\"{a}cklund self-transformation of the KdV6. If an `old' solution $z$ of equation \eqref{e5} satisfies the potential Korteweg--de~Vries equation \eqref{e22}, then the `new' solution $u$ of equation \eqref{e5}, related to $z$ by transformation \eqref{e26}--\eqref{e27} with notations \eqref{e25}, is also a solution of equation \eqref{e22} or satisfies the condition $u_x = \tfrac{1}{2} \sigma$. One can prove this statement (where, of course, $u$ and $z$ are interchangeable) by direct elimination of $z$ from the system of equations \eqref{e26}, \eqref{e27} and $z_t + z_{xxx} + 6 z_x^2 = 0$, the result being $( 2 u_x - \sigma ) \left( u_t + u_{xxx} + 6 u_x^2 \right) = 0$. For example, if we apply the B\"{a}cklund transformation \eqref{e26}--\eqref{e27} to the solution $z = 0$ of equation \eqref{e5}, which is also a solution of equation \eqref{e22}, we get
\begin{multline} \label{e28}
	u = 2 \rho \Bigl( c_1 \exp ( 2 \rho x ) + c_2 \exp \left( 8 \rho^3 t \right) \Bigr)^2 \biggl( c_1^2 \exp ( 4 \rho x ) - c_2^2 \exp \left( 16 \rho^3 t \right) \\ + 4 \rho \Bigl( c_1 c_2 \left( x - 12 \rho^2 t \right) + c_3 \Bigr) \exp \left( 2 \rho x + 8 \rho^3 t \right) \biggr)^{-1} ,
\end{multline}
where $\rho$, $c_1$, $c_2$ and $c_3$ are arbitrary constants, $\rho^2 = 2 \sigma$. Of course, solution \eqref{e28} of equation \eqref{e5} turns out to be a solution of equation \eqref{e22} too. Consequently, in order to obtain, using the B\"{a}cklund transformation \eqref{e26}--\eqref{e27}, a nontrivial solution of equation \eqref{e5} which is not a solution of equation \eqref{e22}, one must apply the transformation to a solution of equation \eqref{e5} which is not a solution of equation \eqref{e22} either, i.e.\ one must initially know some solution of the KdV6 which does not satisfy the potential Korteweg--de~Vries equation.

\section{Solutions and symmetries} \label{s4}

Let us try to find the `wave of translation' solution
\begin{equation} \label{e29}
	u = U ( X ) , \qquad X = x - s t , \qquad s = \text{constant}
\end{equation}
of the KdV6, assuming that $U$ is asymptotically constant at $X \to \pm \infty$. From equation \eqref{e5} we get the sixth-order nonlinear ordinary differential equation
\begin{equation} \label{e30}
	U^{(\text{vi})} + ( 20 U' - s ) U^{(\text{iv})} + 40 U'' U ''' + 12 ( 10 U' - s ) U' U'' = 0 .
\end{equation}
Then we use the substitution
\begin{equation} \label{e31}
	Y ( X ) = - U' + \tfrac{1}{20} s ,
\end{equation}
integrate equation \eqref{e30} over $X$, and get the fourth-order equation
\begin{equation} \label{e32}
	Y^{(\text{iv})} = 20 Y Y'' + 10 {Y'}^2 - 40 Y^3 + \tfrac{3}{10} s^2 Y - \tfrac{1}{100} s^3 ,
\end{equation}
where we have fixed the constant of integration in accordance with the required asymptotic condition
\begin{equation} \label{e33}
	Y \to \tfrac{1}{20} s \quad \text{at} \quad X \to \pm \infty .
\end{equation}

Equation \eqref{e32} is a special case, with $\kappa = 0$, $\alpha = \tfrac{3}{10} s^2$ and $\beta = - \tfrac{1}{100} s^3$, of the Cosgrove's F-V equation
\begin{equation} \label{e34}
	Y^{(\text{iv})} = 20 Y Y'' + 10 {Y'}^2 - 40 Y^3 + \alpha Y + \kappa X +\beta
\end{equation}
possessing the Painlev\'{e} property \cite{C}. Let us remind that, by the method used in \cite{C}, the F-V equation \eqref{e34} with $\kappa = 0$ can be solved in terms of hyperelliptic functions as follows:
\begin{equation} \label{e35}
	Y ( X ) = \tfrac{1}{4} \bigl( \mu ( X ) + \nu ( X ) \bigr) ,
\end{equation}
where the functions $\mu$ and $\nu$ are determined by the equations
\begin{equation} \label{e36}
	I_1 ( \mu ) + I_1 ( \nu ) = K_3 , \qquad I_2 ( \mu ) + I_2 ( \nu ) = X + K_4 ,
\end{equation}
$K_3$ and $K_4$ are arbitrary constants, $I_1$ and $I_2$ are defined as
\begin{equation} \label{e37}
	I_1 ( \mu ) = \int^\mu \! \! \dfrac{d \tau}{\sqrt{P ( \tau )}} \, , \qquad I_2 ( \mu ) = \int^\mu \! \! \dfrac{\tau d \tau}{\sqrt{P ( \tau )}} \, ,
\end{equation}
and in the quintic polynomial
\begin{equation} \label{e38}
	P ( \tau ) = \tau^5 - 2 \alpha \tau^3 + 8 \beta \tau^2 + 32 K_1 \tau + 16 K_2
\end{equation}
the arbitrary constants $K_1$ and $K_2$ correspond to the integrals
\begin{equation} \label{e39}
	Y' H' - 4 Y J - \tfrac{1}{2} H^2 - \left( 2 Y^2 - \tfrac{1}{4} \alpha \right) H = K_1 , \qquad {H'}^2 - 8 H J = K_2
\end{equation}
of equation \eqref{e34} with $\kappa = 0$, the notations being
\begin{equation} \label{e40}
	H = Y'' - 6 Y^2 + \tfrac{1}{4} \alpha , \qquad J = Y Y'' - \tfrac{1}{2} {Y'}^2 - 4 Y^3 + \tfrac{1}{4} \beta .
\end{equation}

In our case \eqref{e32} of the F-V equation we get $K_1 = \tfrac{3}{1000} s^4$ and $K_2 = \tfrac{9}{6250} s^5$ from the asymptotic condition \eqref{e33} and relations \eqref{e39}--\eqref{e40}, and the quintic polynomial \eqref{e38} takes the form
\begin{equation} \label{e41}
	P ( \tau ) = \left( \tau - \tfrac{3}{5} s \right)^2 \left( \tau + \tfrac{2}{5} s \right)^3 .
\end{equation}
This allows us to compute integrals \eqref{e37} in terms of inverse hyperbolic functions and obtain from relations \eqref{e35}--\eqref{e36} the following solution of equation \eqref{e32}:
\begin{equation} \label{e42}
	Y = \dfrac{\xi^2}{10} \left( 2 - \dfrac{5 \left( 1 + 2 \xi^2 ( X - \eta )^2 - \cosh ( 2 \xi X ) \right)}{\bigl( \sinh ( \xi X ) - \xi ( X - \eta ) \cosh ( \xi X ) \bigr)^2} \right) ,
\end{equation}
where the constant $\xi$ is determined by $s = 4 \xi^2$, the arbitrary constant $\eta$ is related to $K_3$, whereas the arbitrary constant $K_4$ has been fixed by an appropriate shift of $x$ in $X$. Finally, relations \eqref{e29}, \eqref{e31} and \eqref{e42} lead us to the following quite simple travelling wave solution of equation \eqref{e5}:
\begin{equation} \label{e43}
	u = \left( \dfrac{1}{\xi \tanh ( \xi X )} - \dfrac{1}{\xi^2 ( X - \eta )} \right)^{-1} ,
\end{equation}
where $\xi$ and $\eta$ are arbitrary constants, $\xi \neq 0$, $X = x - 4 \xi^2 t$, and the additive constant of integration has been omitted in the right-hand side for convenience. One can generalize this solution \eqref{e43} by arbitrary constant shifts of $x$ and $u$, $x \mapsto x + \zeta_1$ and $u \mapsto u + \zeta_2$.

\begin{figure}
	\includegraphics[width=6cm]{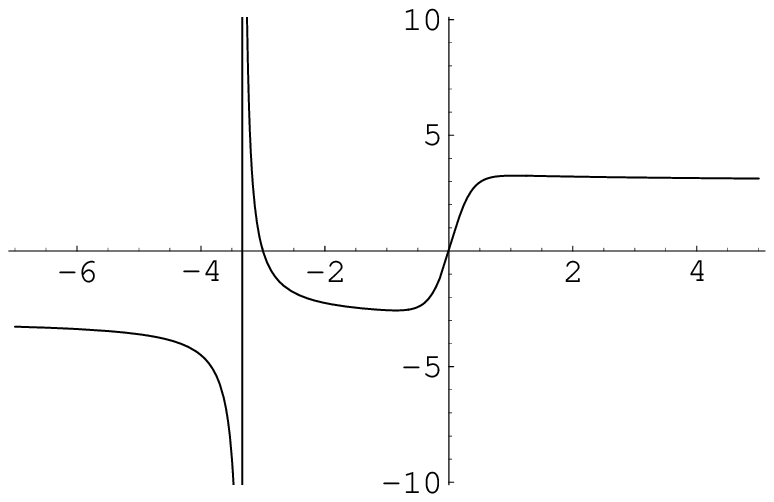}
	\hfil
	\includegraphics[width=6cm]{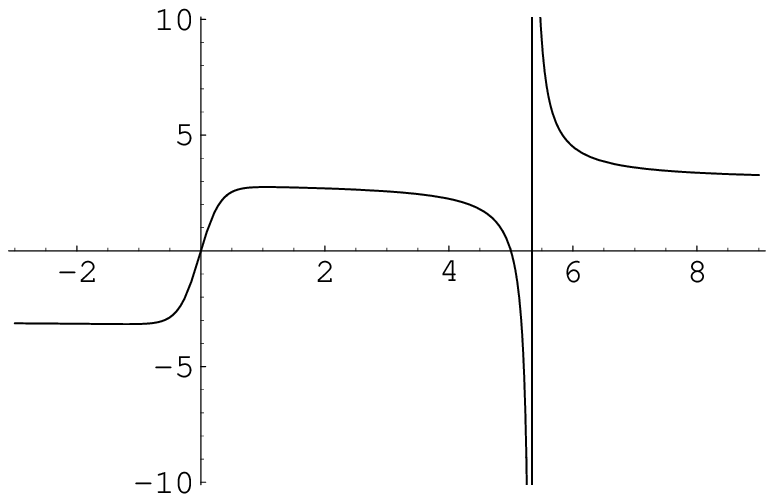}
	\caption{The travelling wave solution \eqref{e43} of equation \eqref{e5}, $u(X)$ with $\xi = 3$: $\eta = -3$ (left) and $\eta = 5$ (right).} \label{fig1}
\end{figure}

It is easy to guess from figure~\ref{fig1} that the obtained solution \eqref{e43} is a nonlinear superposition of two elementary travelling wave solutions of equation \eqref{e5},
\begin{equation} \label{e44}
	u = \xi \tanh \left( \xi x - 4 \xi^3 t \right)
\end{equation}
and
\begin{equation} \label{e45}
	u = \xi \left( \pm 1 - \left( \xi x - 4 \xi^3 t \right)^{-1} \right)^{-1} ,
\end{equation}
where $\xi$ is an arbitrary constant, $\xi \neq 0$ in expression \eqref{e45}, and that the arbitrary constant $\eta$ determines the `distance' between these two solutions in their superposition \eqref{e43}. We do not know, however, why solutions \eqref{e44} and \eqref{e45} do not appear separately when we solve equation \eqref{e32} by the method used in \cite{C}. Let us also note that solution \eqref{e44} of equation \eqref{e5} is the well-known one-soliton solution of equation \eqref{e22}, whereas solutions \eqref{e43} and \eqref{e45} of the KdV6 do not satisfy the potential Korteweg--de~Vries equation.

Integrable nonlinear wave equations usually possess infinitely many generalized symmetries \cite{O}. Using the Dimsym program \cite{D} based on the standard prolongation technique for computing symmetries, we found the following three third-order generalized symmetries of equation \eqref{e5}:
\begin{gather}
	S_1 = \left( u_{xxx} + 6 u_x^2 \right) \partial_u , \label{e46} \\
	S_2 = \left( 3 t u_{xxx} + 18 t u_x^2 - x u_x - u \right) \partial_u , \label{e47} \\
	S_3 = h (t) \left( u_t + u_{xxx} + 6 u_x^2 \right) \partial_u , \label{e48}
\end{gather}
where $h (t)$ is an arbitrary function. On available computers, however, we were unable to find any symmetry of order higher than three. In variables \eqref{e13} the obtained symmetries \eqref{e46}--\eqref{e48} correspond to the Lie point symmetries
\begin{gather}
	\bar{S}_1 = ( w_x - v_t ) \partial_v + 4 ( v w_x - w v_x ) \partial_w , \label{e49} \\
	\bar{S}_2 = \bigl( 3 t (w_x - v_t ) - x v_x -2 v \bigr) \partial_v + \bigl( 12 t ( v w_x - w v_x ) - x w_x - w \bigr) \partial_w , \label{e50} \\
	\bar{S}_3 = h (t) w_x \partial_v + \bigl( h (t) ( 4 v w_x - 4 w v_x + w_t ) + h' (t) w \bigr) \partial_w \label{e51}
\end{gather}
of system \eqref{e12}, respectively. We notice that symmetry \eqref{e46} is also a generalized symmetry of the potential Korteweg--de~Vries equation \eqref{e22}. This allows us to guess, and then to prove by direct computation in accordance with the definition of recursion operators \cite{O}, that the well-known recursion operator
\begin{equation} \label{e52}
	R = \partial_x^2 + 8 u_x - 4 \partial_x^{-1} \cdot u_{xx}
\end{equation}
of equation \eqref{e22} is also a recursion operator of equation \eqref{e5}. Consequently, the KdV6 possesses at least one infinite hierarchy of generalized symmetries, of the form $R^n u_x$ ($n = 1, 2, \dotsc$), where $R$ is the recursion operator \eqref{e52}. We believe, however, that the complete algebra of generalized symmetries of the KdV6 can be more rich and interesting.

\section{Conclusion} \label{s5}

In this paper we discovered the new integrable nonlinear wave equation \eqref{e5}, which we call the KdV6, and obtained first results on its properties.

We discovered the KdV6 by applying the Painlev\'{e} test for integrability to the multiparameter class of equations \eqref{e1}. This result provides additional empirical confirmation of sufficiency of the Painlev\'{e} property for integrability. Then we derived the Lax pair \eqref{e20}--\eqref{e21} and B\"{a}cklund self-transformation \eqref{e26}--\eqref{e27} of equation \eqref{e5}, using the method of truncated singular expansion. We found that the KdV6 is associated with the same spectral problem as of the potential Korteweg--de~Vries equation \eqref{e22}, and observed an interesting property of the obtained B\"{a}cklund transformation concerning solutions of equation \eqref{e5} which satisfy equation \eqref{e22}. Finally, we derived the `wave of translation' solution \eqref{e43} of the KdV6, which is more general than the soliton solution of the potential Korteweg--de~Vries equation, and also found the third-order generalized symmetries \eqref{e46}--\eqref{e48} of equation \eqref{e5} and its recursion operator \eqref{e52}.

Taking into account the obtained results, we believe that the KdV6 deserves further investigation. The following problems seem to be interesting:
\begin{itemize}
	\item What is the multisoliton solution of the KdV6? Do solutions of the rational type \eqref{e45} interact elastically with each other and with solutions of the soliton type \eqref{e44}?
	\item Is there any other B\"{a}cklund self-transformation of the KdV6, different from the one obtained in this paper, which can produce nontrivial solutions not satisfying equation \eqref{e22}, when applied to solutions of equation \eqref{e22}?
	\item Is there any other recursion operator of the KdV6, different from the recursion operator \eqref{e52} of the potential Korteweg--de~Vries equation?
	\item What are Hamiltonian structures and conservation laws of the KdV6?
\end{itemize}
And, of course, to find any applications of the new integrable nonlinear wave equation \eqref{e5} to physics and technology is also an interesting problem.

\section*{Acknowledgements}

S.S. is grateful to the Scientific and Technical Research Council of Turkey (T\"{U}B\.{I}TAK) for support and to the Middle East Technical University (ODT\"{U}) for hospitality.

\end{document}